\documentclass[11pt]{article}
\usepackage[margin=1in]{geometry}
\usepackage{amsmath}
\usepackage{graphicx}
\usepackage{caption}
\usepackage{subcaption}

\def\beq{\begin{equation}}
\def\eeq#1{\label{#1}\end{equation}}
\def\eeqn{\end{equation}}

\def\beqa{\begin{eqnarray}}
\def\eeqa#1{\label{#1}\end{eqnarray}}
\def\eeqan{\end{eqnarray}}

\let\bar=\overbar

\def\Dslash{\not{\hbox{\kern-4pt $D$}}}
\def\dslash{\not{\hbox{\kern-2pt $\del$}}}

\def\msb{{\bar{\ssstyle M \kern -1pt S}}}

\def\Title#1{\begin{center} {\Large {\bf #1} } \end{center}}
\def\Author#1{\begin{center} {\normalsize {\sc #1} } \end{center}}
\def\Institution#1{\begin{center} {\normalsize {\it #1} } \end{center}}
\def\Abstract#1{\noindent {\normalsize {\bf Abstract:} {\normalfont #1}}}
\def\Conference{\vspace{4mm}\begin{raggedright} {\normalsize {\it Talk presented at the 2019 Meeting of the Division of Particles and Fields of the American Physical Society (DPF2019), July 29--August 2, 2019, Northeastern University, Boston, C1907293.} } \end{raggedright}\vspace{4mm}}

\begin{document}

%
%

\Title{Status on the Search for $K_L^0 \rightarrow \pi^0 \nu \bar{\nu}$ with the KOTO Experiment}

\Author{Melissa A. Hutcheson}
\normalsize{\centering{on behalf of the KOTO Collaboration \\}}
\Institution{Department of Physics \\ University of Michigan, Ann Arbor, MI, USA}

\Abstract{The KOTO experiment at the J-PARC research facility in Tokai, Japan aims to observe and measure the rare decay of the neutral kaon, $K_L^0 \rightarrow \pi^0 \nu \bar{\nu}$. This decay has a Standard Model (SM) predicted branching ratio (BR) of $(3.00 \pm 0.30) \times 10^{-11}$~\cite{theoryPaper}. While this decay is extremely rare, it is one of the best decays in the quark sector to probe for new physics beyond the SM due to small theoretical uncertainties. The E391a experiment at KEK 12-GeV PS previously searched for $K_L^0 \rightarrow \pi^0 \nu \bar{\nu}$ events and set a limit of BR($K_L^0 \rightarrow \pi^0 \nu \bar{\nu}$) $< 2.6 \times 10^{-8}$ in 2010~\cite{E391aResults}. In 2018, KOTO set a new experimental limit of BR($K_L^0 \rightarrow \pi^0 \nu \bar{\nu}$) $< 3.0 \times 10^{-9}$ from data collected in 2015, improving the best experimental upper limit by an order of magnitude~\cite{2015Paper}. From 2016 to 2018, KOTO collected around 1.5 times more data than in 2015, and the analysis of this dataset is  underway. This report covers the progress and current status of the 2016-2018 data analysis.}

\Conference

%
%

\section{Introduction and Motivation}
\subsection{Neutral Kaon Decay}
\begin{figure}[htb]
\centering
\includegraphics[height=1.7in]{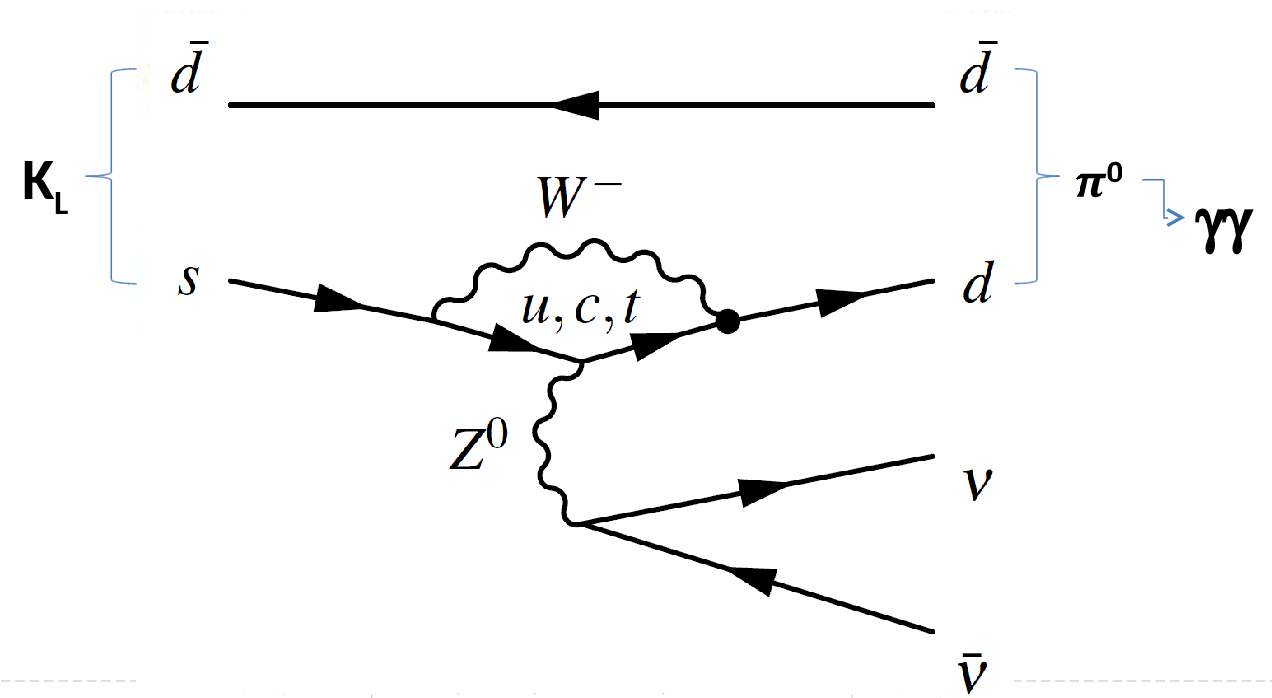}
\caption{SM diagram of $K_L^0 \rightarrow \pi^0 \nu \bar{\nu}$ in which either a u, c, or t quark mediates the decay of an s quark to a d quark through a loop process. Image courtesy of \cite{celestethesis}.}
\label{fig:feynmandiagram1}
\end{figure}

The matter and antimatter asymmetry that we see in the universe today can be partially explained by CP (charge-parity) violation in particle interactions and decays, though we do not fully understand the extent to which matter dominates. Understanding CP violation is one of the central goals of particle physics and there have been many experiments since its discovery that have continued to investigate this phenomenon in both the quark and lepton sectors. In particular, the golden decay, $K_L^0 \rightarrow \pi^0 \nu \bar{\nu}$, is a good probe to test for new physics beyond the SM. This decay directly violates CP and is a Flavor Changing Neutral Current (FCNC) process that proceeds through second order weak interactions as shown in Figure~\ref{fig:feynmandiagram1}. \\

\begin{figure}[htb]
\centering
\includegraphics[height=1.9in]{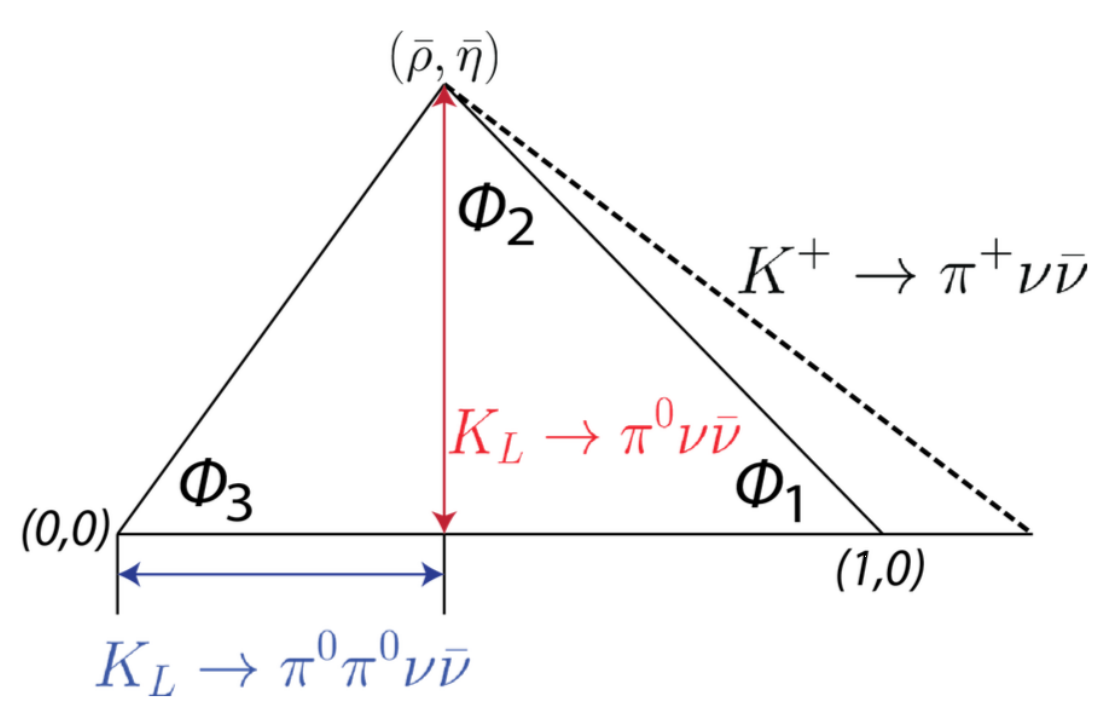}
\caption{The kaon unitary triangle. The dotted line demonstrates potential new physics contributions. Image courtesy of \cite{celestethesis}.}
\label{fig:unitarytriangle}
\end{figure}

\noindent
The broken symmetry of CP is explained by the Cabibbo-Kobayashi-Maskawa (CKM) model \cite{ckmpaper1} \cite{ckmpaper2}, and the decay amplitude of $K_L^0 \rightarrow \pi^0 \nu \bar{\nu}$ is proportional to the imaginary part of a product of CKM matrix elements. In addition, a unitary triangle can be formed in which the branching ratio of $K_L^0 \rightarrow \pi^0 \nu \bar{\nu}$ is proportional to the height of the unitary triangle (Figure~\ref{fig:unitarytriangle}).  

\noindent
The SM predicts the BR of $K_L^0 \rightarrow \pi^0 \nu \bar{\nu}$ to be $(3.00 \pm 0.30) \times 10^{-11}$  with theoretical uncertainties around 1-2\% \cite{theoryPaper}. The current best experimental upper limit is $< 3.0 \times 10^{-9}$ at the 90\% C.L. which was set by KOTO in 2019, from data that was collected in 2015 \cite{2015Paper}. This improved the previous upper limit set by the E391a collaboration \cite{E391aResults} by an order of magnitude. The KOTO experiment expects to reach a Single Event Sensitivity (SES) of $3 \times 10^{-11}$ \cite{kotoproposal}. In doing so, the results of the KOTO experiment will either establish more precise limits for the SM or lead to new physics beyond the Standard Model (BSM). In addition, the KOTO experiment will be able to constrain the height of the unitary triangle which will be a critical check for new physics. The study of CP violation through this decay mode will continue in the future with a proposed experiment to measure $K_L^0 \rightarrow \pi^0 \nu \bar{\nu}$ called KLEVER beginning in 2025 at CERN \cite{KLEVERProposal}.

\subsection{Charged Kaon Decay}
In addition to measuring the neutral kaon decay mode, much effort has gone into measuring the branching fraction of $K^+ \rightarrow \pi^+ \nu \bar{\nu}$. In fact, by using the branching ratio of the charged kaon decay, we can set an indirect limit on BR($K_L^0 \rightarrow \pi^0 \nu \bar{\nu}$). This model-independent upper bound on the branching ratio called the Grossman-Nir bound \cite{grossmannir}, is derived from isospin symmetry arguments as

\begin{center}
 BR($K_L^0 \rightarrow \pi^0 \nu \bar{\nu}$) $< 4.4 \times$ BR($K^+ \rightarrow \pi^+ \nu \bar{\nu}$).
 \end{center}

\begin{figure}[ht]
\begin{center}
\includegraphics[width=0.8\columnwidth]{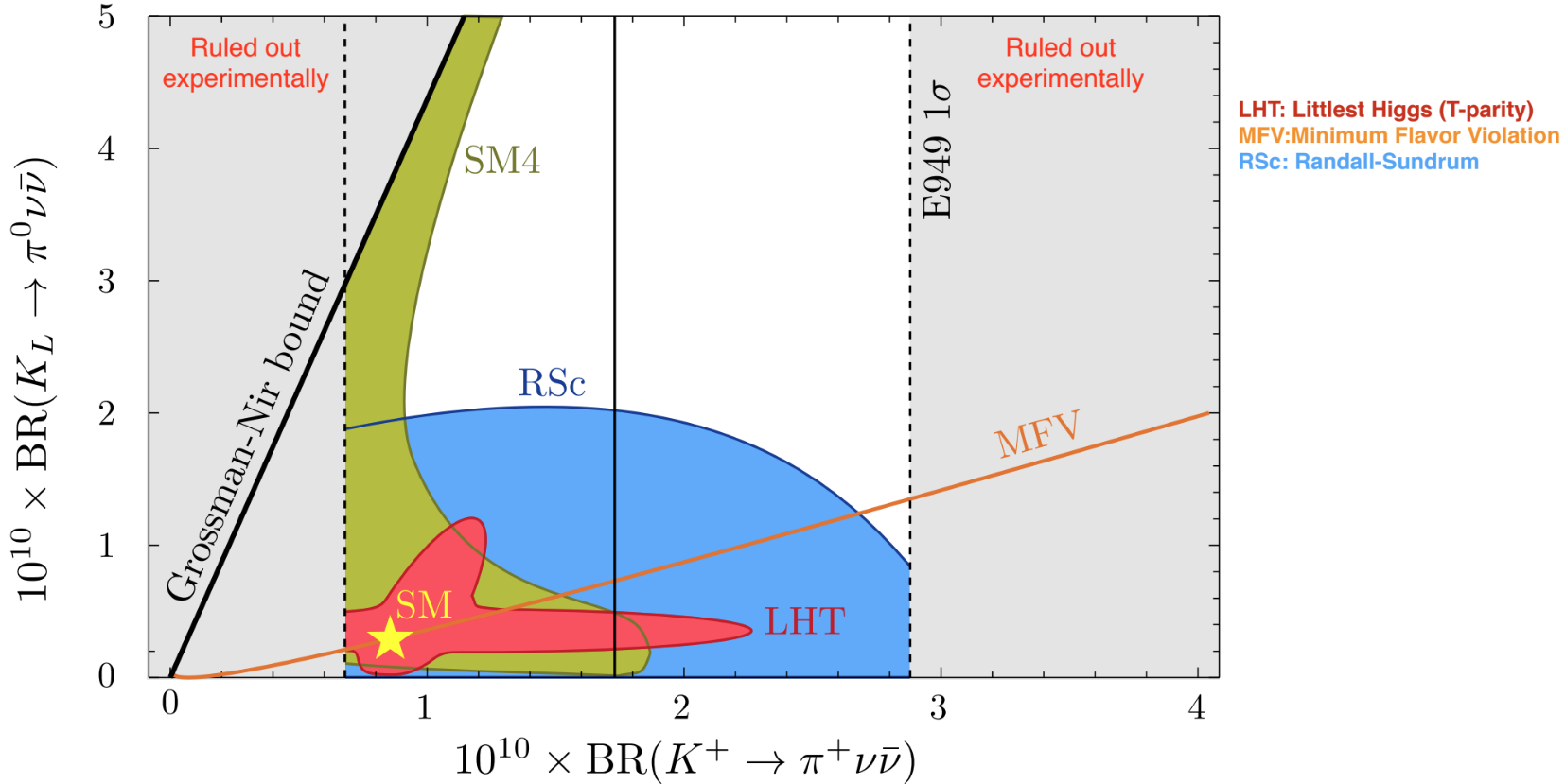}
\end{center}
\caption{Branching ratios for $K_L^0 \rightarrow \pi^0 \nu \bar{\nu}$ and $K^+ \rightarrow \pi^+ \nu \bar{\nu}$ for various models. The E949 experiment measurement is shown in black with $1\sigma$ bounds. Image courtesy of \cite{GrossmanNirPlot}. 
\label{fig:GNSUSY}}
\end{figure}

\noindent
Thus, the measured BR($K^+ \rightarrow \pi^+ \nu \bar{\nu}$) of $(1.7 \pm 1.1) \times 10^{-10}$ from the BNL E949 experiment \cite{E949paper} gives an upper limit on BR($K_L^0 \rightarrow \pi^0 \nu \bar{\nu}$) to be $1.5 \times 10^{-9}$ at the 90\% C.L. The NA62 experiment at CERN is a continuation of the E949 experiment and aims to measure BR($K^+ \rightarrow \pi^+ \nu \bar{\nu}$) which has a SM predicted BR of $(9.11 \pm 0.72) \times 10^{-11}$ \cite{theoryPaper}. Recent results from NA62 report one observed event which gives an upper limit of BR($K^+ \rightarrow \pi^+ \nu \bar{\nu}$) $ < 14 \times 10^{-10}$ at the 95\% C.L. \cite{NA62Results}. Figure \ref{fig:GNSUSY} shows the BR possibilities for both $K_L^0 \rightarrow \pi^0 \nu \bar{\nu}$ and $K^+ \rightarrow \pi^+ \nu \bar{\nu}$ for various models. \\

\newpage
\noindent
While the charged decay mode has been relatively well constrained, the neutral decay mode still has a large range of possibilities for the branching ratio. A combination of both $K_L^0 \rightarrow \pi^0 \nu \bar{\nu}$ and $K^+ \rightarrow \pi^+ \nu \bar{\nu}$ measurements is necessary to developing a better understanding of the underlying CP violating processes. \\

\section{Experimental Method}
\begin{figure}[ht]
\begin{center}
\includegraphics[width=1.0\columnwidth]{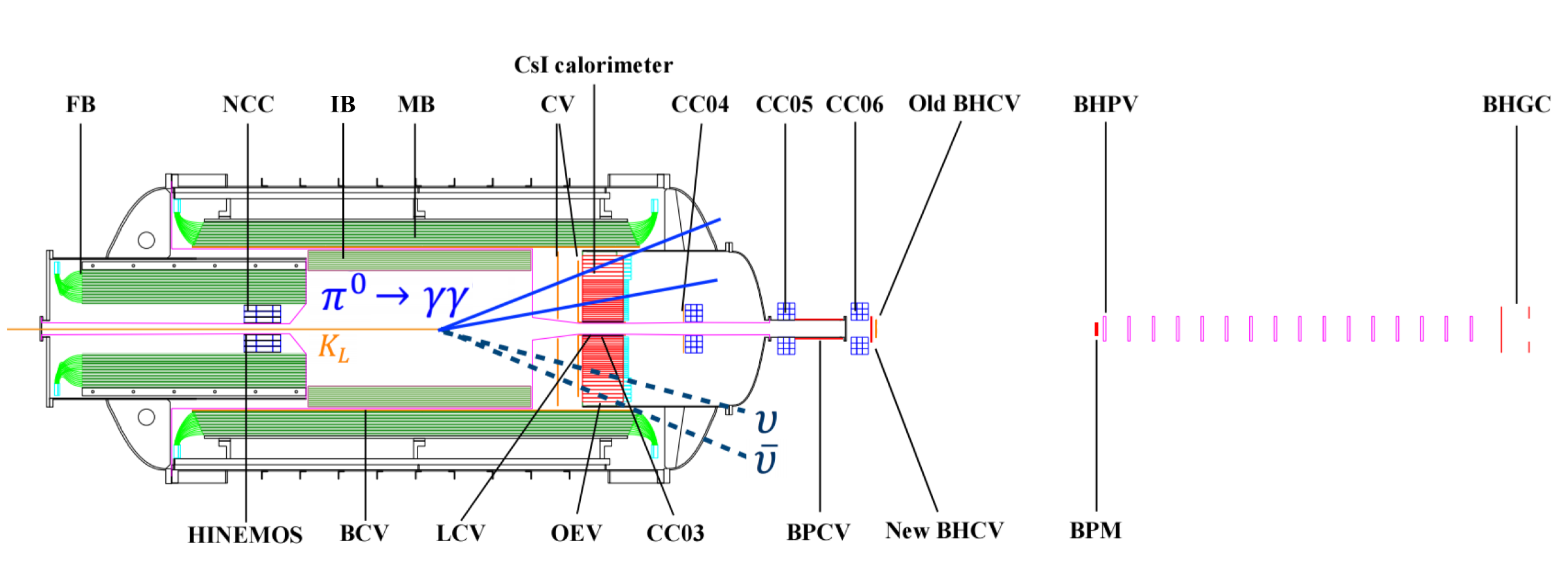}
\end{center}
\caption{Sideview of the KOTO detectors with the $K_L^0 \rightarrow \pi^0 \nu \bar{\nu}$ decay signal. Image modified from \cite{stephpaper}. \label{fig:detector}}
\end{figure}

The KOTO experiment located at the J-PARC research facility in Tokai, Japan is a fixed target experiment in which a 30 GeV beam of protons collides with a stationary gold target. Using collimators and magnets to sweep away charged particles, a neutral "pencil" beam of kaons is produced. The KOTO detectors are located 21.5 meters from the production target and kaons from the secondary beam decay within the vacuum region. The KOTO detectors consist of a hermetic system of scintillating veto detectors and a Cesium Iodide (CsI) calorimeter (Figure \ref{fig:detector}). The signal decay is two photons from the pion which hit the CsI calorimeter and the two neutrinos are not seen by the KOTO detectors. The experimental difficulty in observing $K_L^0 \rightarrow \pi^0 \nu \bar{\nu}$ is due to the lack of charged particles in the final state and due to the high efficiency required for the detection of extra particles from other decay modes, such as $K_L^0 \rightarrow 3\pi^0$, $K_L^0 \rightarrow 2\pi^0$, and $K_L^0 \rightarrow \pi^0 \pi^+ \pi^-$. The strategy of KOTO is to observe the two photons from the $\pi^0$ decay with no other particles present in the detectors, in which the two photons have a discernible, large, transverse momentum.

\section{Data Collection History}
\begin{figure}[ht]
\begin{center}
\includegraphics[width=0.7\columnwidth]{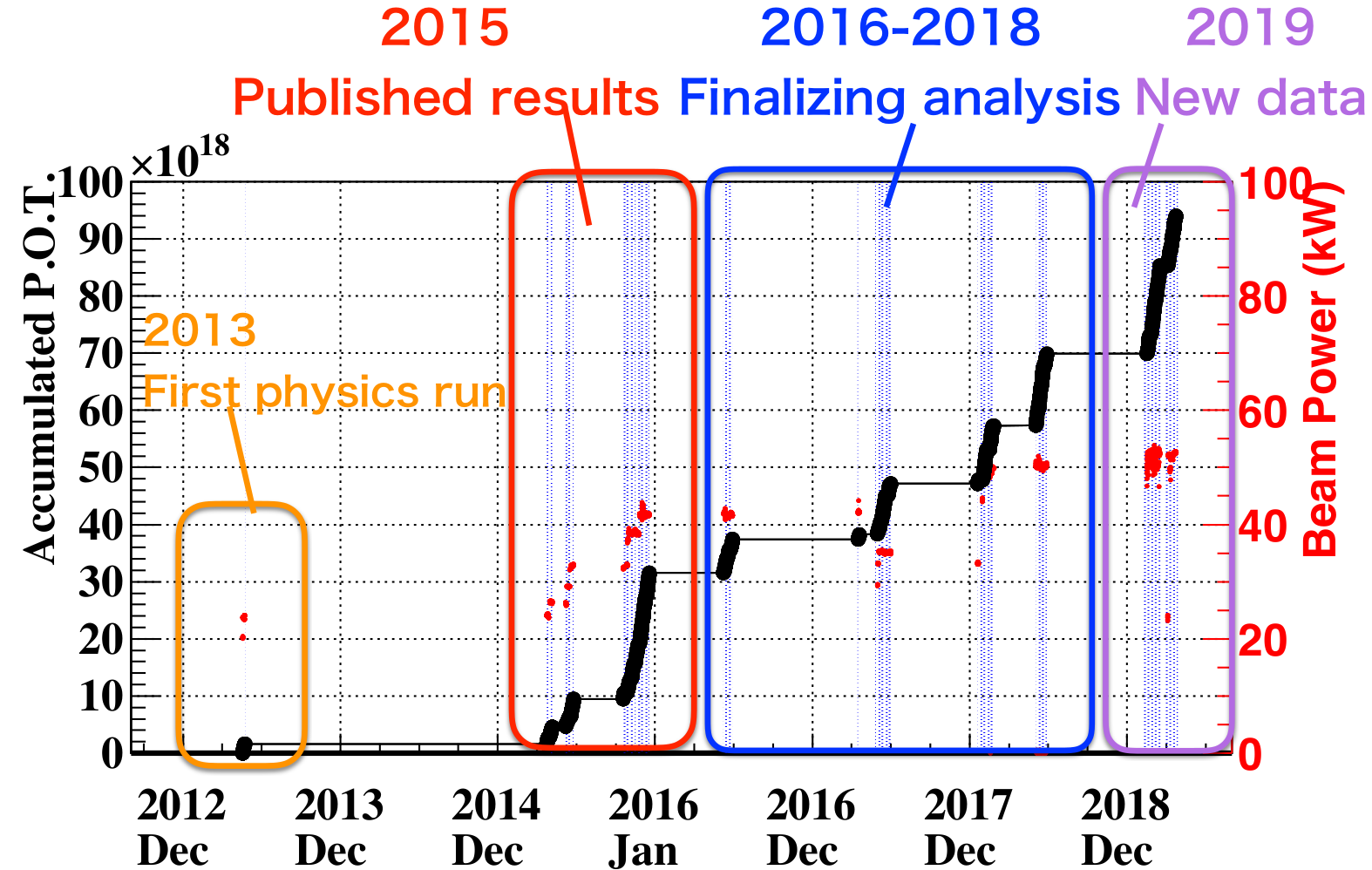}
\end{center}
\caption{The accumulated protons on target (POT) over KOTO's running history. \label{fig:POTshiomi}}
\end{figure}

KOTO took its first 100 hours of physics data in 2013 and the results of this data give an upper limit of BR($K_L^0 \rightarrow \pi^0 \nu \bar{\nu}$) $< 5.1 \times 10^{-8}$ at the 90\% C.L. \cite{2013runpaper}.  KOTO began running again in 2015 and since then has been taking data runs periodically as shown in Figure \ref{fig:POTshiomi}. The accumulated protons on target (POT) gives an indication of how much data has been collected, along with the status of each data set. From the 2015 data analysis, a new best experimental upper limit of BR($K_L^0 \rightarrow \pi^0 \nu \bar{\nu}$) $< 3.0 \times 10^{-9}$ at the 90\% C.L. was set \cite{2015Paper}. Following the 2015 runs, a new detector was installed and improvements on the trigger and data analysis techniques were made. The details of these upgrades are described at the end of this section. From 2016 to 2018 KOTO collected around 1.5$\times$ more data than in 2015 with a total accumulated POT of $3.1 \times 10^{19}$. The analysis status of this data set is detailed in the following sections. 

\subsection{Upgrades and Improvements after 2015}
In the summer of 2015, the detector was upgraded with the installation of the inner barrel (IB) to reduce the $K_L^0 \rightarrow 2\pi^0$ background. The IB consists of 25 layers of 5 mm scintillator and 24 layers of 1 mm lead plates. This detector was inserted inside the main barrel and added an additional 5 radiation lengths to prevent photon punch through in the main barrel. The estimated suppression of the $K_L^0 \rightarrow 2\pi^0$ background is roughly 30\%. The data acquisition (DAQ) system was also upgraded in 2017 to include cluster finding in the trigger to improve the DAQ efficiency. The new trigger is based on energy instead of ADC counts which allows us to select events of interest efficiently. This trigger calculates cluster information for each event allowing us to increase the number of $K_L^0 \rightarrow 3\pi^0$ and $K_L^0 \rightarrow 2\pi^0$ events which are needed for the normalization analysis, and eliminate events that are not used in analysis. \\

\noindent
In the 2015 runs, we implemented a new technique to study the largest background seen in 2013, the hadron cluster background. This background comes from halo neutrons in the beam that produce hadronic showers in the CsI calorimeter and fake a signal event with two clusters. To study this background of neutrons, we took special runs to collect a control sample in which a movable aluminum plate was inserted into the beamline to scatter neutrons. From 2016 to 2018 we collected 8 times more data of the control sample to improve the discrimination between neutrons and photons. In addition, new cuts and algorithms were developed to improve the discrimination between neutrons and photons. In particular, the performance of the cluster shape discrimination method was improved with a convolutional neural network which uses deep-learning and takes inputs based on the energy and timing of the crystals in the cluster. The pulse shape discrimination method was also improved by using a Fourier transformation of the raw waveforms which extracted the characteristics more clearly.

\section{Analysis}
The principle challenge of KOTO is the reduction of background from other $K_L^0$ decay modes as well as hadronic shower events in the CsI due to neutron interactions in the detectors. Because of this, a thorough analysis of backgrounds is critical in order to observe the signal decay. This is done through detailed Monte Carlo studies and a blind analysis method that makes cuts before the data in the signal region is revealed. There are three variables needed in order to calculate the branching ratio (Equation \ref{eq:BRcalculation}): the number of reconstructed $K_L^0 \rightarrow \pi^0 \nu \bar{\nu}$ events ($N_\text{signal}$), the number of $K_L^0$ generated at the beam exit ($N_\text{$K_L^0$}$), as well as the acceptance of the signal mode ($A_\text{signal}$). 

\begin{equation}
BR(K_L^0 \rightarrow \pi^0 \nu \bar{\nu}) = \dfrac{N_\text{signal}}{N_\text{$K_L^0$} \times A_\text{signal}} \label{eq:BRcalculation}
\end{equation}

\noindent
Because we don't know the number of signal events until we unblind the data, it is useful to refer to the Single Event Sensitivity (SES),

\begin{equation}
SES = \dfrac{1}{N_\text{$K_L^0$} \times A_\text{signal}}. \label{eq:SES}
\end{equation}

\noindent
The smaller the SES, the higher the chance we are able to observe a signal event. Therefore, we want as many as possible $K_L^0$s generated and a high acceptance. The next few sections will detail how these variables are determined and cover various background reduction methods. 

\subsection{Signal Reconstruction}
\begin{figure}[ht]
\centering
\begin{minipage}{0.45\textwidth}
  \centering
  \includegraphics[width=1\linewidth]{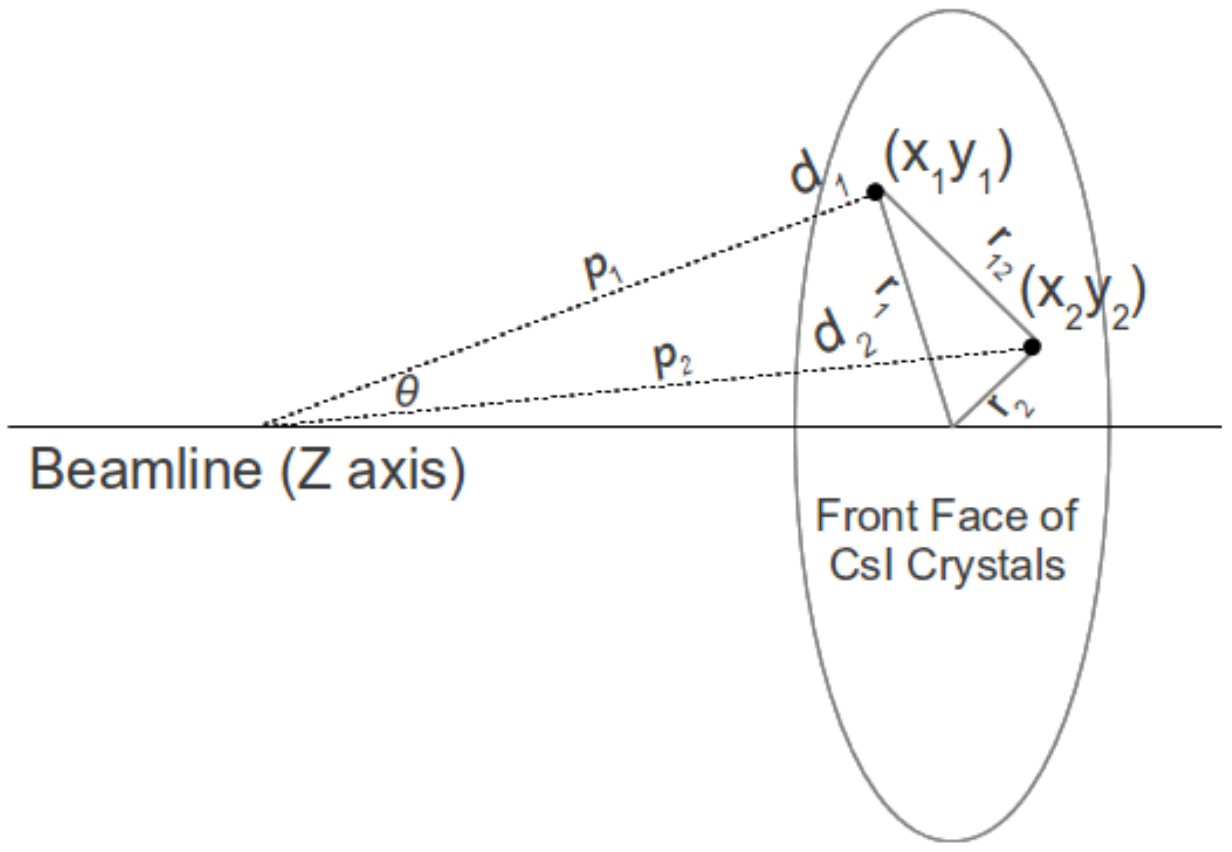}
  \caption{$\pi_0$ reconstruction from two photons. Image courtesy of \cite{celestethesis}.}
  \label{fig:reconstruction}
\end{minipage}%
\hspace{1cm}
\begin{minipage}{0.45\textwidth}
  \centering
  \includegraphics[width=1\linewidth]{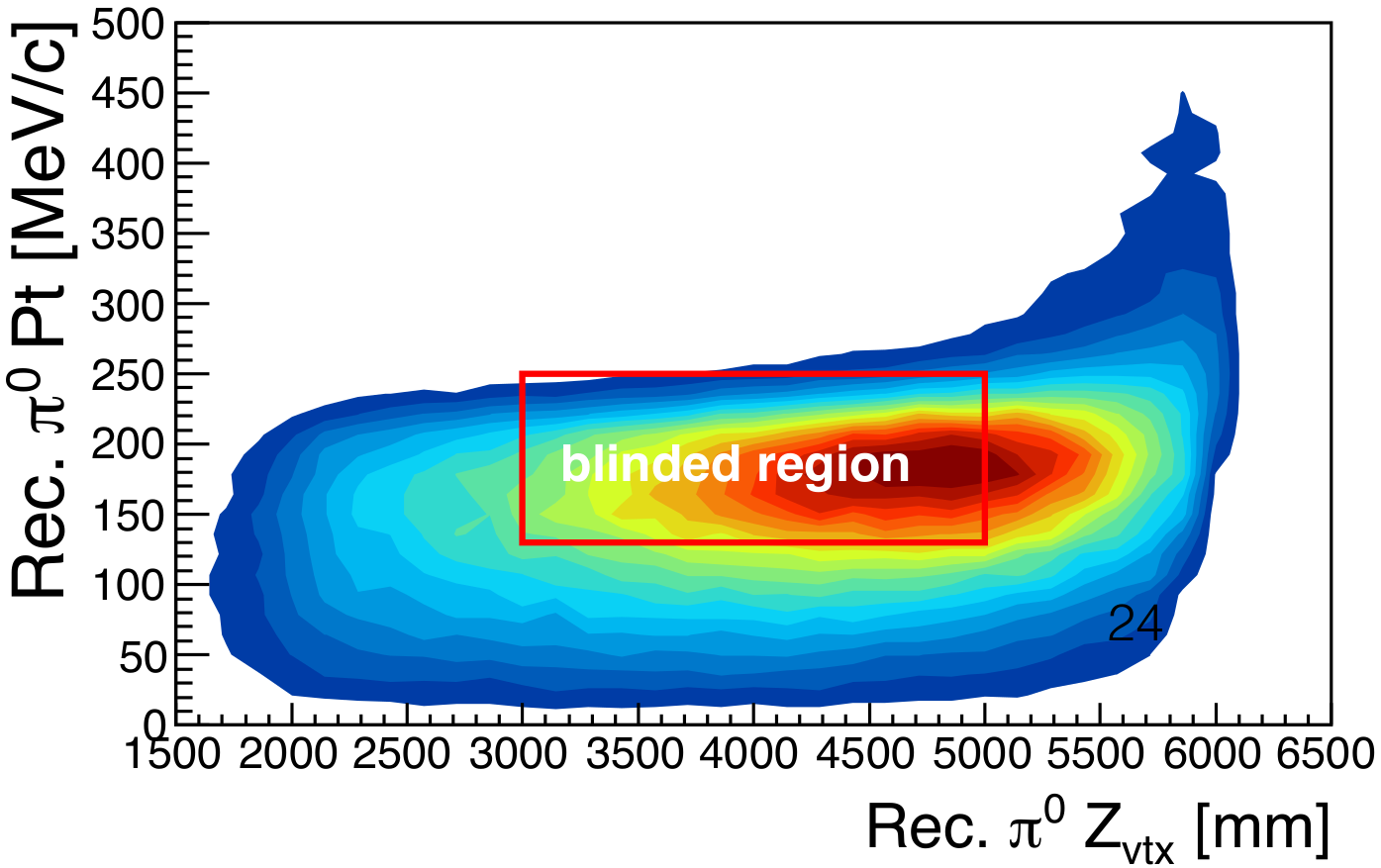}
  \caption{Monte Carlo sample of $K_L^0 \rightarrow \pi^0 \nu \bar{\nu}$ signal distribution.}
  \label{fig:MonteCarloSignal}
\end{minipage}
\end{figure}

In order to identify $K_L^0 \rightarrow \pi^0 \nu \bar{\nu}$ events and calculate $N_\text{signal}$,  the decay vertex of the pion ($\pi^0 \rightarrow \gamma \gamma$) must be reconstructed. From the CsI calorimeter we only know the position and energy of the two photons, not the incident angles at which they hit the CsI. In order to do a kinematic reconstruction, we use the mass of the $\pi^0$ as a constraint as well as constrain the decay position to be on the beamline (Figure \ref{fig:reconstruction}). Events with only two cluster hits on the CsI and no signals in the veto detectors are selected. From this, the decay vertex (Z position) and transverse momentum ($P_t$) can be reconstructed. Figure \ref{fig:MonteCarloSignal} shows a Monte Carlo sample of the $K_L^0 \rightarrow \pi^0 \nu \bar{\nu}$ signal distribution from the reconstructed $P_t$ and Z vertex. 

\subsection{Normalization}
Calculating the number of $K_L^0$ events generated at the beam exit requires knowing what the flux of $K_L^0$s is into the detector. The decay modes used to calculate the $K_L^0$ flux are $K_L^0 \rightarrow 3\pi^0$, $K_L^0 \rightarrow 2\pi^0$, and $K_L^0 \rightarrow 2\gamma$. These normalization modes have large branching fractions and therefore large statistics to accurately calculate the $K_L^0$ flux. The flux is defined as, 

\begin{equation}
K_L^0 \text{ flux} = \dfrac{K_L^0 \text{ yield}}{\text{POT}_\text{runs} / \text{POT}_\text{norm factor}} \label{eq:Flux}
\end{equation}

\noindent
where $\text{POT}_\text{runs}$ is the number of protons on target (POT) scaled by the normalization prescale. The $K_L^0 \text{ yield}$ is calculated as 

\begin{equation}
K_L^0 \text{ yield} = \dfrac{N_\text{data}}{A_\text{total}} \label{eq:Yield}
\end{equation}

\noindent
where $N_\text{data}$ is the number of events in data remaining after applying all kinematic and veto cuts, and $A_\text{total}$ is the total acceptance which is a sum of the acceptance for each mode multiplied by the branching ratio for that mode. The yield and flux are calculated separately for each mode and then multiplied by the total number of POT to get the number of $K_L^0$s generated at the beam exit: 

\begin{equation}
N_\text{$K_L^0$} = \dfrac{K_L^0 \text{ flux}}{\text{POT}_\text{norm factor}} \times \text{POT} \label{eq:NumberofKLs}
\end{equation}

\noindent
The $K_L^0 \rightarrow 2\pi^0$ mode is used in the final evaluation of the total number of $K_L^0$s because it has an energy profile closest to the energy profile of the signal. The total number of $K_L^0$s calculated for the 2016-2018 data set is $(7.14 \pm 0.05) \times 10^{12}$. This is 1.57$\times$ more than the number of $K_L^0$s collected in 2015. \\

\begin{figure}
\begin{subfigure}{.5\textwidth}
  \centering
  \includegraphics[width=.65\linewidth]{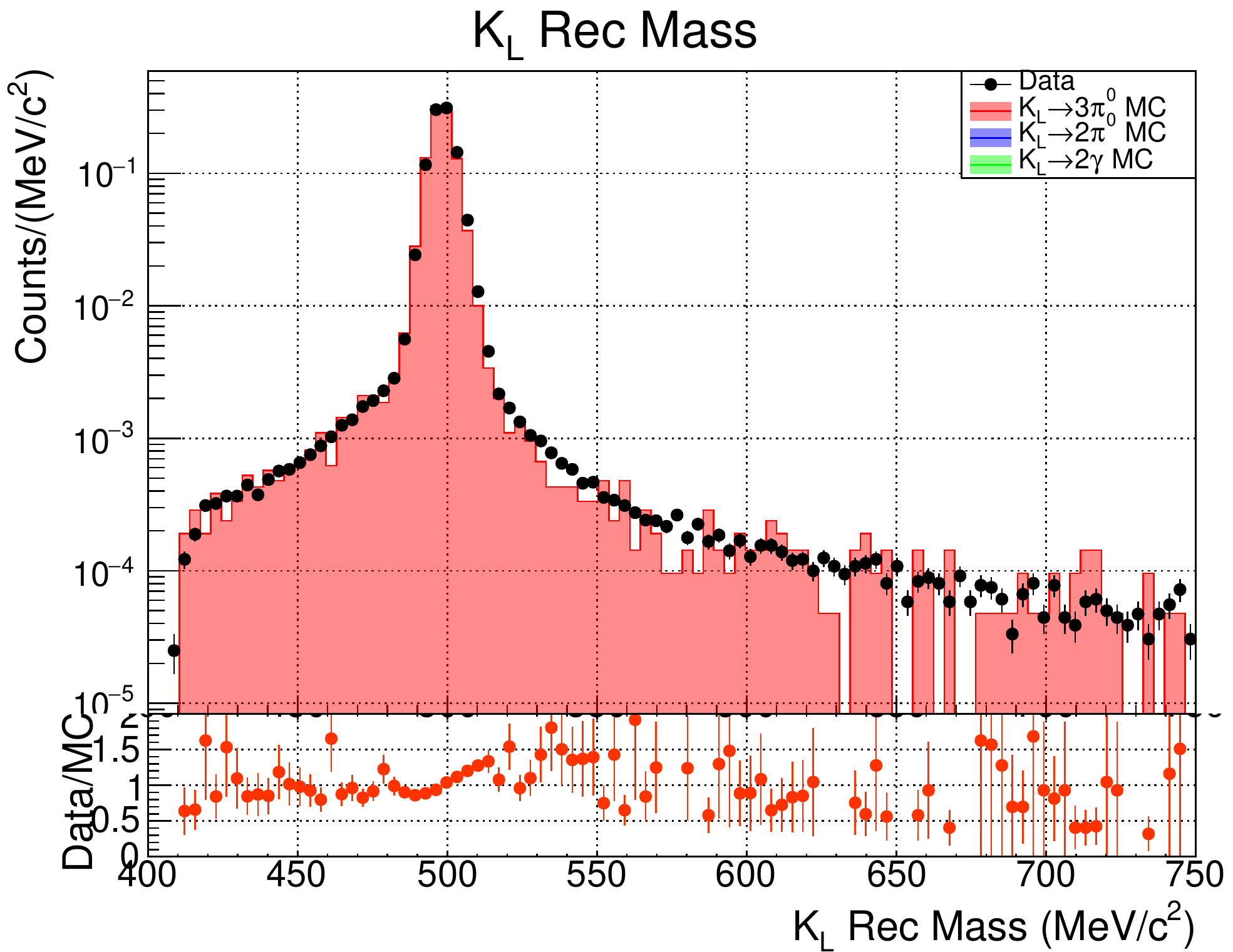}  
  \caption{$K_L^0 \rightarrow 3\pi^0$ reconstructed $K_L$ mass}
  \label{fig:3pi0KLRecMass}
\end{subfigure}
\begin{subfigure}{.5\textwidth}
  \centering
  \includegraphics[width=.65\linewidth]{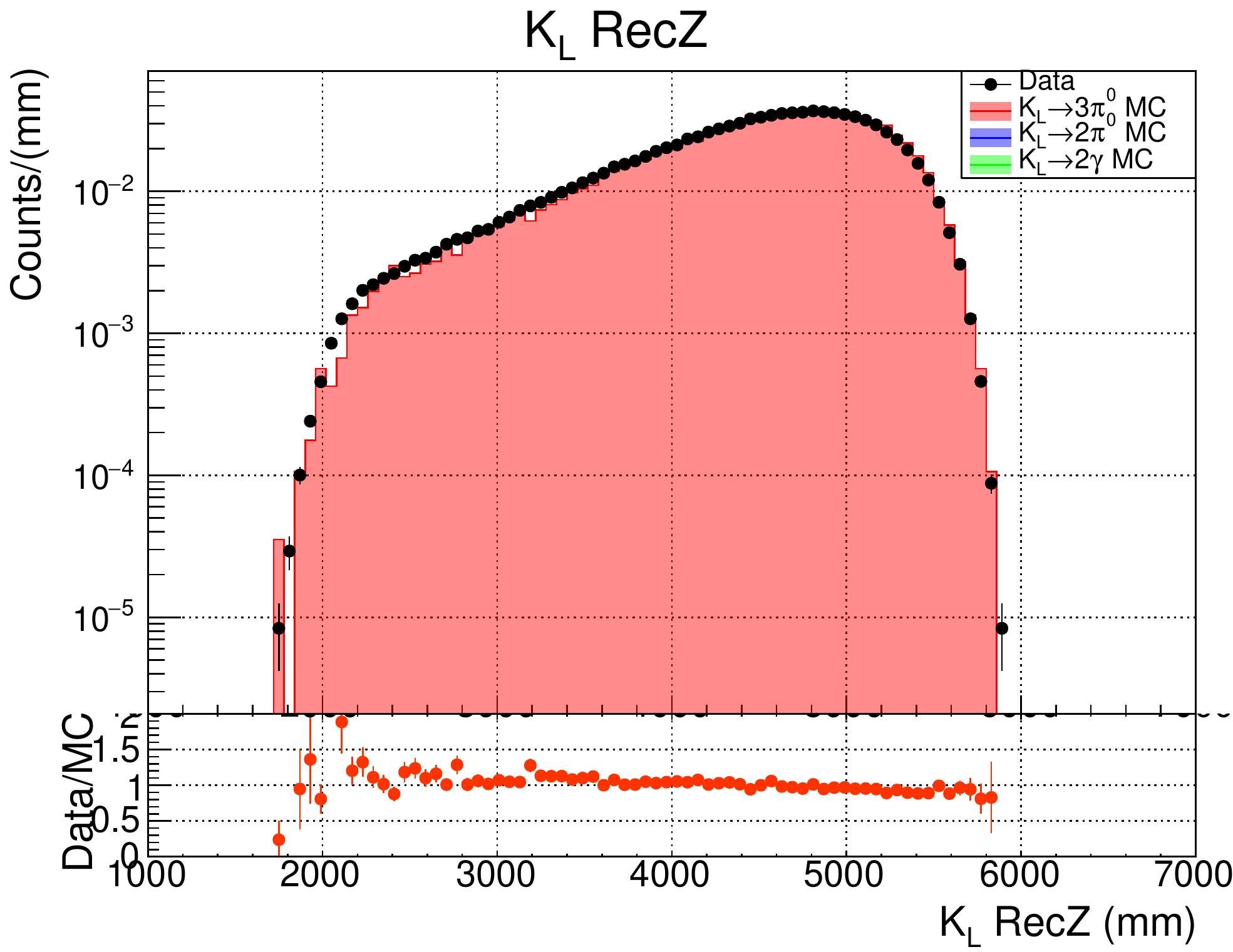}  
  \caption{$K_L^0 \rightarrow 3\pi^0$ reconstructed $K_L$ Z vertex}
  \label{fig:3pi0KLRecZ}
\end{subfigure}


\begin{subfigure}{.5\textwidth}
  \centering
  \includegraphics[width=.65\linewidth]{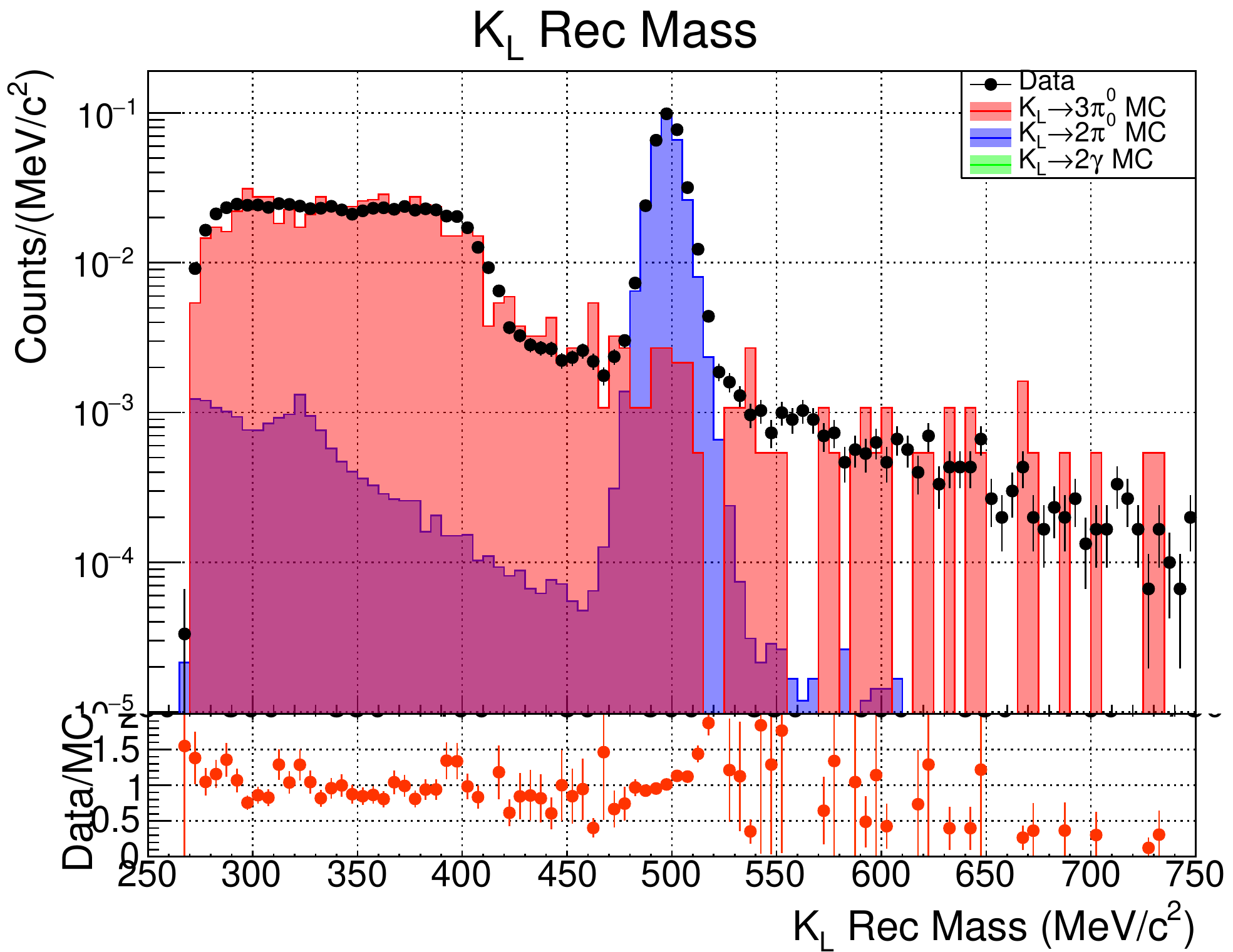}  
  \caption{$K_L^0 \rightarrow 2\pi^0$ reconstructed $K_L$ mass}
  \label{fig:2pi0KLRecMass}
\end{subfigure}
\begin{subfigure}{.5\textwidth}
  \centering
  \includegraphics[width=.65\linewidth]{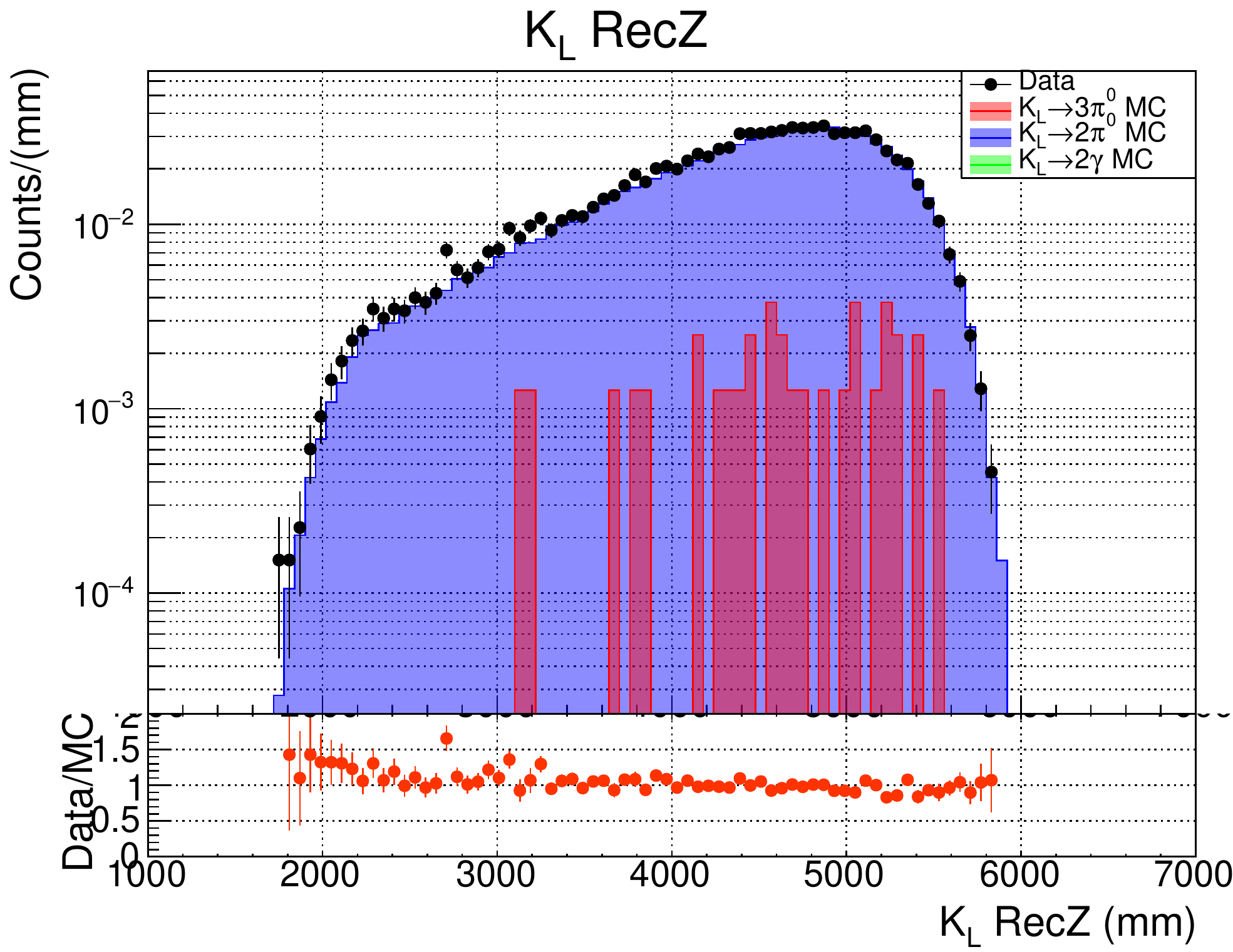}  
  \caption{$K_L^0 \rightarrow 2\pi^0$ reconstructed $K_L$ Z vertex}
  \label{fig:2pi0KLRecZ}
\end{subfigure}


\begin{subfigure}{.5\textwidth}
  \centering
  \includegraphics[width=.65\linewidth]{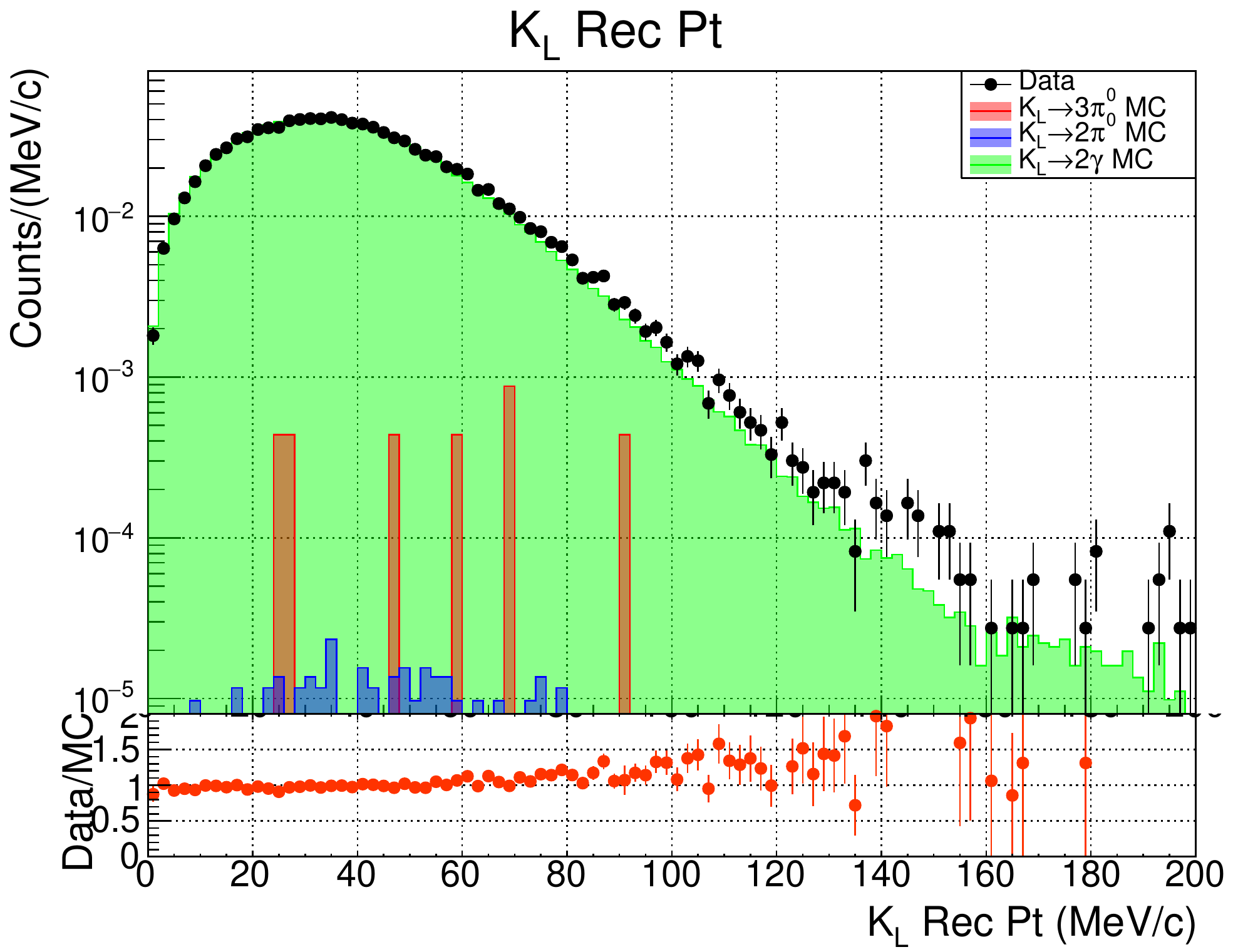}  
  \caption{$K_L^0 \rightarrow 2\gamma$ reconstructed $K_L$ $P_t$}
  \label{fig:2gammaKLRecPt}
\end{subfigure}
\begin{subfigure}{.5\textwidth}
  \centering
  \includegraphics[width=.65\linewidth]{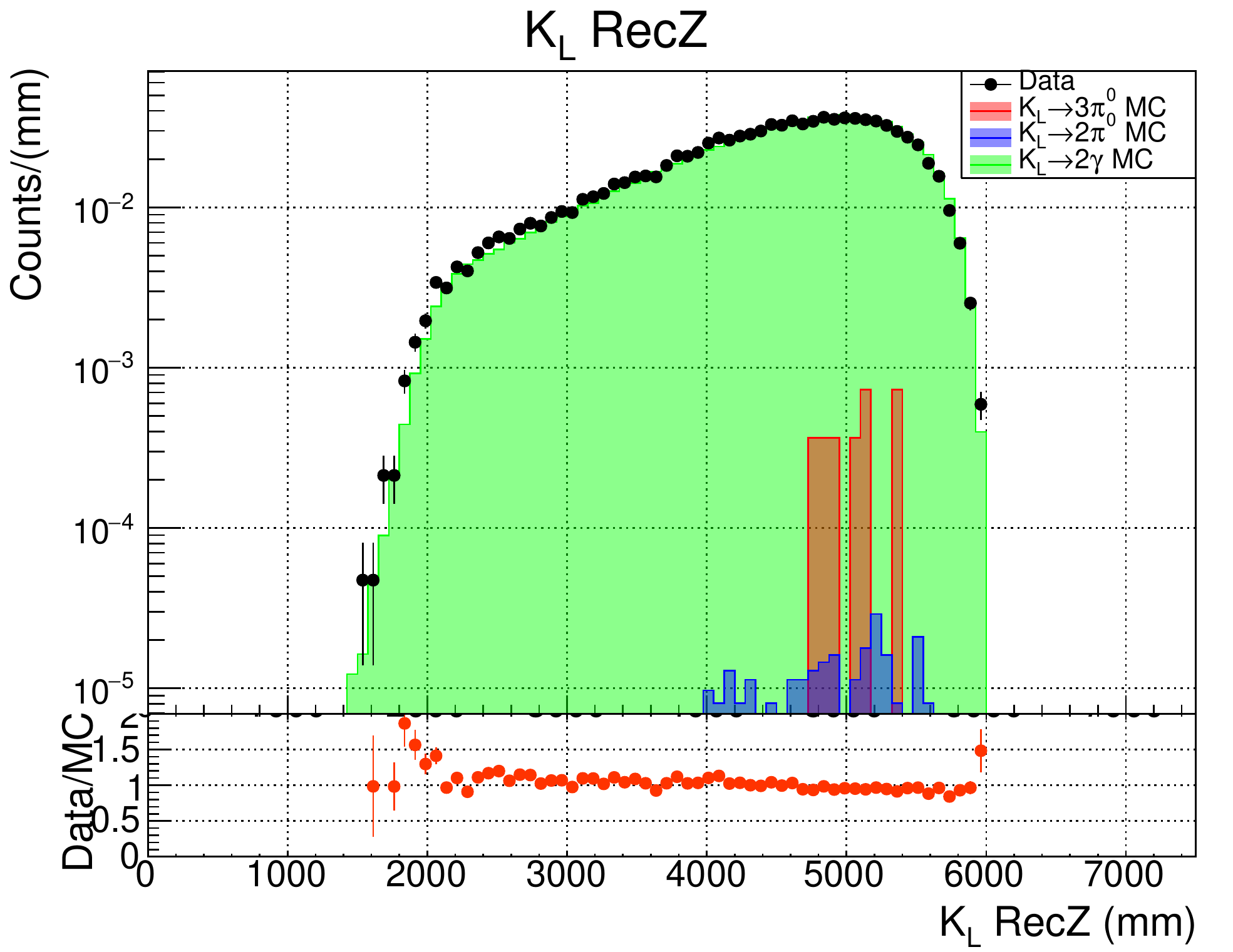}  
  \caption{$K_L^0 \rightarrow 2\gamma$ reconstructed $K_L$ Z vertex}
  \label{fig:2gammaKLRecZ}
\end{subfigure}
\caption{Kinematic distributions for $K_L^0 \rightarrow 3\pi^0$ (red), $K_L^0 \rightarrow 2\pi^0$ (blue), and $K_L^0 \rightarrow 2\gamma$ (green).}
\label{fig:KinematicDists}
\end{figure}

\noindent
Along with calculating the number of $K_L^0$s generated at the beam exit, these modes are also used to evaluate the kinematic and veto cut efficiencies and check the Monte Carlo reproducibility of the data. Figure \ref{fig:KinematicDists} shows kinematic distributions for each normalization mode and there is good agreement between data and Monte Carlo. The last variable in the branching ratio calculation, the signal acceptance ($A_\text{signal}$), includes the geometric acceptance of the detectors and the kinematic and veto efficiencies for the cuts made on the data. 

\subsection{Background Estimation}
While charged veto detectors are in place to reject events with charged particles in the final state, the remaining background still dominates over the signal because of its extremely small branching ratio. Hermetic photon vetos are in place to detect other neutral $K_L^0$ background modes, those of which include $K_L^0 \rightarrow 3\pi^0$, $K_L^0 \rightarrow 2\pi^0$, $K_L^0 \rightarrow \pi^0 \pi^+ \pi^-$, and $K_L^0 \rightarrow \gamma \gamma$. In addition, accidental activities that are off timing may cause an accidental hit on a veto that masks the true veto hit. This masking background comes from $K_L^0 \rightarrow \pi^+ e^- \nu_e$ and $K_L^0 \rightarrow 3\pi^0$. Effective kinematic and veto cuts require careful studies of these background sources, to first estimate them, and then reduce them. \\

\noindent
The other large source of background comes from halo neutrons in the beamline. Some of these neutrons hit the detector material close to the beamline and produce other particles such as a $\pi_0$ or $\eta$ that can fake a signal event. An neutron can also hit the calorimeter directly and create a hadronic shower which produces a second hadronic shower from a neutron in the first shower. Several new detectors were added after the 2013 run in order to reduce the background from these sources. In addition, special analysis reduction methods were developed such as the cluster shape discrimination and pulse shape discrimination methods described in section 3.1. \\

\begin{figure}[ht]
\begin{center}
\includegraphics[width=0.6\columnwidth]{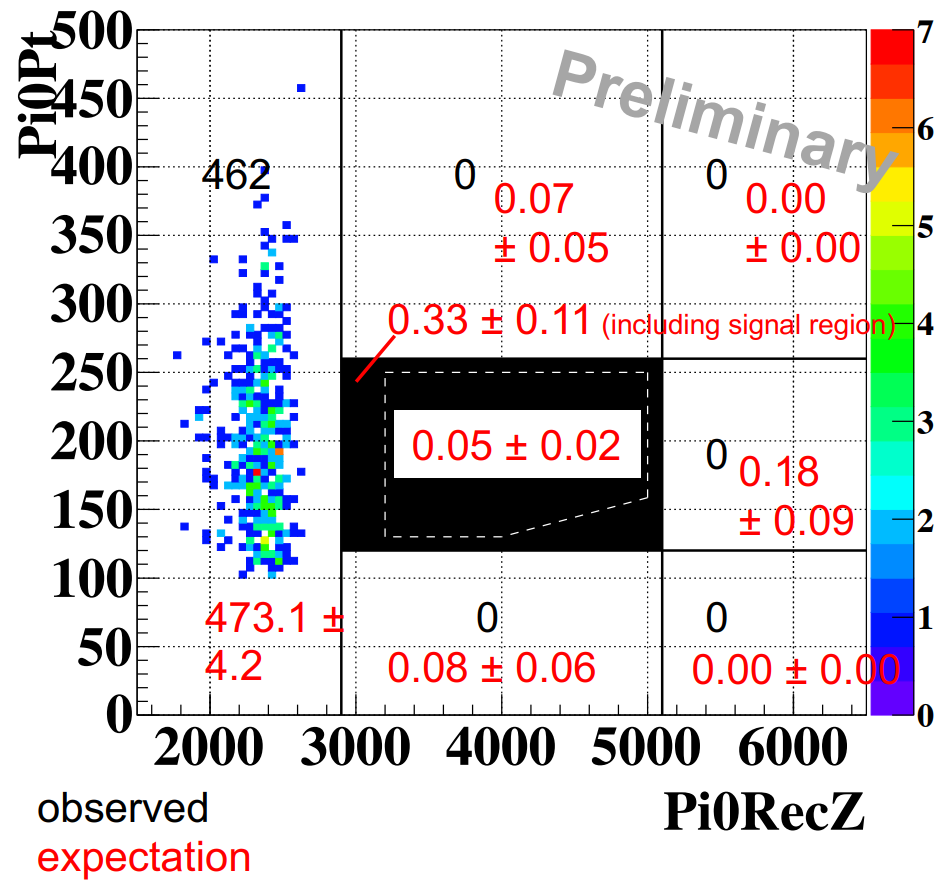}
\end{center}
\caption{$\pi^0$ reconstructed $P_t$ and $Z$ for the 2016-2018 data set. The blinded region is in black and the signal region is the dotted region. The values in red are the expected number of background in each region and the values in black are the observed number of events. \label{fig:PtZBlinded}}
\end{figure}

\noindent
The preliminary 2016-2018 data and background estimations in the $P_t$ and $Z$ plane are shown in Figure \ref{fig:PtZBlinded}. 

\section{Summary}
The KOTO experiment took data from 2016 through 2018 and collected a total of $(7.14 \pm 0.05) \times 10^{12}$ number of $K_L^0$s which is 1.57$\times$ more than the number of $K_L^0$s collected in 2015. The number of protons on target (POT) collected in this data set was $3.1 \times 10^{19}$ and is 1.5$\times$ more than the amount collected in 2015. The estimated Single Event Sensitivity (SES) is $6.9 \times 10^{-10}$. 

\section{Acknowledgements}
This work is supported by the U.S Department of Energy, Office of Science, under research award No.s DE-SC0007859, DE-SC0006497, and DE-SC0009798.

\end{document}